\documentclass[aps,onecolumn,prd,showpacs,showkeys,preprintnumbers,superscriptaddress,nobibnotes,notitlepage,floatfix,longbibliography,nofootinbib,10pt]{revtex4-2}

\usepackage{silence}
\usepackage{booktabs}
\usepackage{multirow}
\usepackage{graphicx}
\usepackage{amsmath}
\usepackage{amssymb}
\usepackage{xcolor}
\usepackage{enumitem}
\usepackage[colorlinks=true,citecolor=blue,urlcolor=blue]{hyperref}

\def\be{\begin{equation}}
\def\ee{\end{equation}}
\def\bea{\begin{eqnarray}}
\def\eea{\end{eqnarray}}

\def\gev{\, {\rm GeV}}

\newcommand{\gsim}{\lower.7ex\hbox{$\;\stackrel{\textstyle>}{\sim}\;$}}
\newcommand{\lsim}{\lower.7ex\hbox{$\;\stackrel{\textstyle<}{\sim}\;$}}

\begin{document}

\title{A likelihood analysis for gamma-ray background models}

\author{Chance Hoskinson}
\affiliation{Department of Physics and Astronomy, University of Utah, Salt Lake City, UT  84112, USA}

\author{Jason Kumar}
\affiliation{Department of Physics and Astronomy, University of Hawai'i, Honolulu, HI 96822, USA}

\author{Pearl Sandick}
\affiliation{Department of Physics and Astronomy, University of Utah, Salt Lake City, UT  84112, USA}

\begin{abstract}
Indirect searches for dark matter using dwarf spheroidal galaxies are limited by systematic uncertainties in modeling diffuse gamma-ray backgrounds. We present a likelihood-based comparison of locally constructed empirical background models and theoretically-motivated models that incorporate the Fermi-LAT diffuse background.  The empirical models we study include both an independent-binning approach and a covariance-based approach that captures cross-energy correlations. Using ensembles of blank-sky regions and information criteria which account for model complexity, we find that empirical background descriptions provide a statistically competitive fit to gamma-ray data on degree scales in high-latitude regions.
\end{abstract}

\maketitle

\section{Introduction}
A promising approach to the search for dark matter (DM) is through indirect detection, in which DM particle annihilation or decay produces Standard Model (SM) particles, such as hadrons, electrons and photons. Photons are especially powerful probes because, unlike charged cosmic rays, they are not deflected by magnetic fields and therefore point back to their sources.\footnote{This is also true of neutrinos, but neutrinos have a small scattering cross section, making them much harder to detect than photons.} This directional information is particularly important when targeting localized gamma-ray sources such as dwarf spheroidal galaxies (dSphs).
Dwarf spheroidal galaxies are believed to be DM-dominated and are expected to produce few gamma rays through standard astrophysical processes~\cite{Strigari:2012acq}, making them excellent laboratories for the indirect detection of dark matter~\cite{Aramaki:2022zpw,Boddy:2022knd}.  These searches are expected to remain important in  the coming years 
as new observatories, including the Vera C. Rubin Observatory, are expected to discover many new dSphs~\cite{LSST:2008ijt,LSSTDarkMatterGroup:2019mwo,Mao:2022fyx}. 

However, in order to search for an excess of gamma rays arriving from the direction of a dSph, it is necessary to correctly model 
the distribution of gamma rays arising from standard astrophysical processes along the line-of-sight to the dSph.  We refer to this 
as modeling of the background, and it is a source of large systematic uncertainty.

The Fermi Large Area Telescope (Fermi-LAT) diffuse model~\cite{FermiModel} is built from theoretical templates of the physical processes that produce gamma rays. This approach carries systematic uncertainties of $\sim 3\%$ in the modeling of diffuse emission, which can be even larger in specific local regions~\cite{Fermi-LAT:2019yla}.  A complementary approach is to create a purely empirical background model from 
gamma rays observed from slightly off-axis from the target~\cite{Geringer_Sameth_2011,Geringer-Sameth:2014qqa}.  The logic of this approach is based on the assumption that diffuse 
gamma ray sources are likely somewhat spatially smooth, allowing one to determine their contribution to flux along the line-of-sight from 
data slightly away from the line-of-sight.
Our goal in this work is to compare how well  background models created through these two approaches 
match the gamma-ray data obtained from a large number of ``blank-sky" regions, which do not contain any known point sources.

To better understand how background modeling affects indirect DM searches, we compare several approaches using a likelihood framework. The models we evaluate are: (i) an independent-binning empirical background model constructed using the method introduced in 
Refs.~\cite{Geringer_Sameth_2011,Geringer-Sameth:2014qqa} and implemented using the recently developed tool \texttt{TweedleDEE} \cite{hoskinson:tweedleDEE2024}, in which photon counts in each energy bin are treated independently; (ii) a multivariate Gaussian model built from the covariance matrix inferred from \texttt{TweedleDEE} photon-count data; and (iii) a background model obtained by using 
the Fermi diffuse emission model~\cite{FermiModel} and the analysis pipeline available in \texttt{Fermitools}~\cite{Fermitools}. 
Models (i) and (ii) 
are purely data-driven and constructed locally from photon counts in off-axis regions which are binned into 16 logarithmically spaced energy bins spanning $1 - 100~\gev$. 
In contrast, model (iii) is theoretically motivated and consists of two physical components: the Galactic diffuse emission and the isotropic background, each derived from global template fits, together with catalog-based point-source templates whose normalizations and spectral parameters are fit to the data~\cite{Fermi-LAT:2019yla,Ballet:2023qzs}.

Here, we randomly select a large number of $1^\circ$ degree sky regions of interest (ROIs) 
which are away from the Galactic Center, the galactic plane, and 
any known gamma-ray point sources.  We then compare the models produced by each approach using the Bayesian Information Criterion (BIC) 
and Akaike Information Criterion (AIC), to see which model is a better description of the data.  Note that although we use three 
approaches to develop background models, each ROI is independent and could be best described by any one of those.  There is no reason why one particular 
approach should produce a background model which better describes the data in every ROI.  For example, in a region in which the astrophysical 
processes which produce gamma rays are well-modeled, a theoretically-driven approach (such approach (iii)) may produce a background 
model which better describes the data than an empirical approach.  But in a different region, where the relevant astrophysical processes have 
some spatial mismodelling, the empirical approach may produce a background model which better describes the data than the theoretical approach.
So rather than concluding whether one approach is better than another, we instead aim to assess in what fraction of the sky each approach produces a background model that is better than another.

The plan of this paper is as follows.  In Section~\ref{sec:EmpiricalModels}, we describe the two empirical approaches to developing a background 
model and in Section~\ref{sec:Fermi} we describe a theoretically-motivated approach. The analysis methodology is presented in Section~\ref{sec:Analysis}, with results following in Section~\ref{sec:Results}.  We  conclude in Section~\ref{sec:Conclusion}.

\section{Purely Empirical Models}
\label{sec:EmpiricalModels}

The central idea behind
empirical background modeling 
is that diffuse foregrounds are spatially extended and smooth (more-so than a point source, such as 
the annihilation of dark matter in a small subhalo). By sampling photon counts from sky regions slightly offset from a target, one can 
empirically estimate the local fore-/background while excluding the target itself. 
We consider 16 equal logarithmically-spaced energy bins in the range $1-100~\gev$.  We take our ROI to be a region of $1^\circ$ opening 
angle.  We then randomly select a large number of sample regions (of the same size as the ROI), drawn from within $10^\circ $ 
of the center of the ROI,   
discarding sample regions which overlap the ROI, which lie within $0.8^\circ$ of an identified 
point source (including point sources within $12^\circ$ of the ROI), or which do not lie entirely within $10^\circ$ of the ROI.  The sample regions which are used to empirically model the background distribution in the ROI  are 
thus independent of the ROI itself, and are uncontaminated by nearby point sources.

\subsection{Independent Energy Bins}

In our first empirical approach, we assume that 
the distribution of photon counts in each of the 16 energy bins are independent of each other.  
We will denote this as model E1.  This method was proposed in Refs.~\cite{Geringer_Sameth_2011,Geringer-Sameth:2014qqa} 
(see also \cite{Boddy:2018qur,Boddy:2019kuw,Boddy:2024tiu}).
In this case, we randomly select $10^5$ sample regions, before discarding those which lie too close to a point source or the ROI.
The probability mass function (PMF) for the photon 
counts from the ROI lying in the $i$th energy bin, $P_i (N_i)$, is then obtained from the normalized histogram of photon counts in that 
energy bin taken over all of the sample regions.  If  $N_i^O$ denotes the observed number of photons in the $i$th energy bin arriving from 
the ROI, then the likelihood of the data is given by
\bea
{\cal L}_\mathrm{E1}(N^O) &=& \prod_i P_i (N_i^O) .
\eea
The PMFs are generated from Fermi-LAT gamma-ray data, processed by \texttt{FermiPy}~\cite{Wood:2017vfl} and \texttt{Fermitools}~\cite{Fermitools} (see Section~\ref{sec:Fermi} for implementation details) using the \texttt{TweedleDEE} software package (for more details, see ~\cite{hoskinson:tweedleDEE2024}).
Note that the $P_i (N_i)$ are generally not Poisson.  This is not surprising, as the flux of gamma rays arriving from 
different sample regions near the ROI can vary due to differences in the astrophysical source processes which occur in different 
sample regions, in addition to Poisson fluctuations.

\subsection{Including Covariance}
In our second empirical approach to background modeling, we consider the possibility that the probability mass functions for different energy 
bins are correlated.
The motivation for this step is that foreground structures, such as gas clouds, can contribute photons across many energies simultaneously. If a target, such as a dSph, happens to lie behind such a structure, it will receive excess photons in multiple energy bins, creating non-trivial correlations between bins. Such cross-energy correlations can mimic or obscure potential dark-matter spectra, and understanding whether independent or covariance-based models describe the local gamma-ray background more accurately could be important.  We 
will denote this background model as E2.

The covariance matrix is constructed from the same empirical photon-count samples used to build the PMFs, and it encodes both the variance within each energy bin and the correlations between bins (i.e., the probability of seeing $N_i$ photons in the $i$th energy bin and $N_j$ photons in the $j$th energy bin). In other words, we need to know the average number of photons in each energy bin and how fluctuations in one bin are related to another, which is formulated as the joint probability distribution, $P_{i,j}(N_i, N_j)$. 
To form the joint probability distribution, we consider $\sim10^6$ randomly sampled regions within $10^\circ$ of the center of the ROI, and 
create a 2D histogram representing the number of sample regions with $N_i$ counts in the $i$th energy bin and $N_j$ counts in the $j$th energy bin.  

The entries of the normalized 2D histogram are $P_{i,j} (N_i, N_j)$, which satisfy $\sum_{N_i, N_j} P_{i,j}(N_i, N_j) = 1$. 

The diagonal elements of the covariance matrix $K$ are given by
\bea
K_{ii} &=& \sum_{N_i} P_i (N_i) \left[N_i - \langle N_i \rangle \right]^2 ,
\eea
where 
\bea
\langle N_i \rangle &=& \sum_{N_i} P_i (N_i) ~ N_i  .
\eea
The off-diagonal elements of $K$ are given by
\bea
K_{ij} &=& \sum_{N_i, N_j} P_{i,j} (N_i, N_j) \left[N_i - \langle N_i \rangle \right] \left[N_j - \langle N_j \rangle \right] .
\eea

We approximate the joint distribution of photon counts across energy bins as a multivariate Gaussian. Under this model, the likelihood is written as:
\bea
{\cal L}_\mathrm{E2} (N^O) &=& (2\pi)^{-N_{bin}/2} \left(\prod_{i=1}^{N_{bin}} \sigma_i^{-1} \right) 
\exp \left[-\frac{1}{2} \sum_{i=1}^{N_{bin}} \sum_{j=1}^{N_{bin}} (N^O_i - \langle N_i \rangle)K^{-1}_{ij}(N^O_j - \langle N_j \rangle) \right] ,
\eea
where $\sigma_i^2$ are the eigenvalues of $K$, $N_{bin}$ is the number of energy bins, and $K^{-1}$ is the inverse of the covariance matrix. In essence, the closer the number of observed counts is to the average counts predicted by the model, the higher the likelihood.

\section{A Theoretically-Motivated Model}
\label{sec:Fermi}

The theoretically-motivated model which we consider is derived from the \texttt{Fermitools} analysis pipeline, and is optimized to maximize the likelihood of the Fermi-LAT data over a $20^\circ \times 20^\circ$ region of the sky 
centered at the ROI (maintaining a $10^\circ$ radius to match the size of models E1 and E2).  
We denote this model as FT.

For each blank-sky region, we analyze data collected by Fermi-LAT from the mission elapsed time range 239557417 -- 681169985 seconds. Events are selected in the energy range $1-100~\gev$, with good time intervals defined by the filters \texttt{DATA\_QUAL > 0} and \texttt{LAT\_CONFIG == 1}. We adopt a zenith-angle cut of $z_{\max} = 90^\circ$ to remove contamination from the Earth limb, following Fermi-LAT recommendations~\cite{FSSC:DataSelection}. We use the Pass 8~\cite{Atwood:2013rka} instrument response function (IRF) \texttt{P8R3\_SOURCE\_V3}. 

The analysis is performed using \texttt{FermiPy}~\cite{Wood:2017vfl} (v1.4.0) and \texttt{Fermitools}~\cite{Fermitools} (v2.4.0). We define the ROI as outlined above, centered on the target position, and adopt a spatial pixel size of $0.05^\circ$. This choice provides sufficient sampling of the LAT point spread function across the full energy range considered. We include the Galactic diffuse emission model \texttt{gll\_iem\_v07}, the isotropic diffuse model \texttt{iso\_P8R3\_SOURCE\_v3\_v1}, and all sources from the 4FGL-DR4 catalog~\cite{Ballet:2023qzs} within a radius of $15^\circ$ from the target.

To stabilize the background model prior to the final likelihood fit, we first perform a staged optimization using \texttt{FermiPy}’s \texttt{optimize()} routine. This method iteratively adjusts parameters to bring them closer to their global likelihood maxima. Concretely, this method first simultaneously fits the normalizations of the brightest components (those contributing the largest fraction of predicted counts), then individually refits the normalizations of the remaining sources (in decreasing predicted count order, skipping very faint ones), and finally fits both shape and normalization parameters for sources with high test statistic from the earlier steps. We then search for additional point sources not present in the initial catalog using \texttt{find\_sources()}, adopting a detection threshold of $\mathrm{TS}\ge 25$, and enforcing a minimum separation of $0.2^\circ$ between new candidates.

% To maintain consistency with previous dSph analysis frameworks~\cite{Fermi-LAT:2013sme, Fermi-LAT:2015att, DiMauro:2022rmq, Song:2024gda, Luque:2023}, we include a point source at the target position modeled with a power-law spectrum. This source acts as a placeholder for potential emission, allowing the diffuse background components to be optimized without bias from unmodeled central flux~\cite{Fermi-LAT:2015att}. 
After updating the model with any newly found sources, we perform a final optimization, following the same \texttt{optimize()} routine described above, before performing the maximum-likelihood fit with \texttt{fit()}. In this final maximum-likelihood fit, we free only the flux normalizations of catalog and newly added sources within $3^\circ$ of the ROI center that satisfy $\mathrm{TS}>20$ (\texttt{free\_sources(pars='norm', distance=3.0, minmax\_ts=[20,None])}). We additionally allow the normalization and photon index of the Galactic diffuse component and the normalization of the isotropic diffuse component (\texttt{galdiff} and \texttt{isodiff}) to vary via \texttt{free\_source()}. 

The optimization and fitting routines were performed over the full $20^\circ \times 20^\circ$ region.  
To compare this model to the empirical models over the $1^\circ$ ROI, we evaluate the expected number of counts from the FT 
model in each energy bin in the central $1^\circ$ ROI after the global optimization. 
We then compute the Poisson likelihood over the $1^\circ$ ROI, using the expression 

\bea
\mathcal{L}_\mathrm{FT} (N_i^O)
= 
\prod_{i=1}^{N_{\rm bin}}
\frac{\mu_i^{\, N_i^O }}{N_i^O !}\,
e^{-\mu_i} ,
\label{eq:poisson_like}
\eea
where $N_i^O$ is the number of observed photon counts for energy bin $i$, and $\mu_i$ is the expected number of counts 
from the FT model.

\section{Background Model Selection}
\label{sec:Analysis}

For any choice of an ROI, we can consider the comparison of the different background models as a model selection problem.  
Our goal is essentially to determine which background model for the ROI is preferred by the observed data from the ROI.
Because these background models are not nested, a simple comparison of data likelihoods is not necessarily sufficient.    
In particular, because model FT is obtained by maximizing the likelihood while varying model parameters, its raw likelihood values cannot be compared directly to those of the empirical models, which contain no free parameters. Allowing parameters to vary will always increase the likelihood, so to account for this we employ the Bayesian Information Criterion (BIC), defined as
\bea
\mathrm{BIC} &=& k \ln n - 2 \ln {\cal L}^{\text{max}} ,
\eea
where $n$ is the number of data points (in our case, the number of energy bins), $k$ is the number of free model parameters, and ${\cal L}^{\text{max}}$ is the maximized likelihood. For the empirical models E1 and E2, $k = 0$, and ${\cal L}^{\text{max}}$ is simply the computed likelihood. For model FT, ${\cal L}^{\text{max}}$ is the Poisson likelihood obtained for Fermi-LAT data in the ROI 
(eq.~\ref{eq:poisson_like}) after optimizing the model over a $20^\circ \times 20^\circ$ region, and $k$ is the number of parameters freed during the fit.
Since larger values of $k$ increase the BIC, smaller values of BIC indicate a preferred model. 
For any two background models M1 and M2, we can compute the quantity
\bea
\Delta \mathrm{BIC} (\mathrm{M1},\mathrm{M2}) &=&  (k_\mathrm{M1} - k_\mathrm{M2}) \ln n - 2 
(\ln {\cal L}_\mathrm{M1}^{\text{max}} - \ln {\cal L}_\mathrm{M2}^{\text{max}} ) .
\eea
If this quantity is negative (positive), that indicates model M1 (M2) provides a better description of the 
Fermi-LAT data from the ROI.  We perform this comparison pair-wise for all three combinations of models, in 
each ROI.
Because both ${\cal L}^{\text{max}}$ and the number of freed parameters vary across ROIs, we analyze the distribution of 
$\Delta \mathrm{BIC}$ values to assess how often the data prefer one of model E1, E2, and FT over the other two.

We also examine the Akaike Information Criterion (AIC), defined as
\bea
\mathrm{AIC} &=& 2k - 2 \ln {\cal L}^{\text{max}} .
\eea
The AIC assigns a weaker penalty for additional parameters. 
One may roughly characterize the difference between the BIC and AIC by noting that the BIC is expected to 
provide a better criterion for model selection when one of the models contains the true model within its parameter 
space, whereas the AIC is expected to provide a better criterion when both models are expected to only provide an 
approximation.  
For completeness, we will consider both the BIC and AIC.  

\section{Results}
\label{sec:Results}
Before presenting the results, we briefly recap the three models compared in each ROI. 
Model E1 is the empirical independent-binning approach, in which the likelihood is the product of per-energy-bin PMFs constructed from nearby off-axis regions. 
Model E2 is the empirical covariance model, where the joint photon count distribution across energy bins is modeled as a multivariate Gaussian inferred from the same off-axis samples. 
Model FT is the theoretically motivated Fermi-LAT template model, in which the Galactic diffuse, isotropic, and catalog source templates are fit (with free parameters) to maximize the Poisson likelihood.

To evaluate the performance and robustness models E1, E2, and FT relative to each other, we randomly generate two sets of 100 blank-sky ROIs. These ROIs are required to be free of extended sources (within the $10^\circ$ ROI) and sufficiently far from the Galactic plane to avoid contamination from structured diffuse emission.
The two sets differed by progressively more conservative cuts:

\begin{enumerate}
    \item Set A (less restrictive)
    \begin{itemize}
        \item Minimum angular separation of $1^\circ$ between the ROI and all known gamma-ray catalog sources.
        \item Galactic latitude cut of $|b| \ge 15^\circ$, which avoids the brightest central regions of the Galaxy while still providing a sample of high-latitude sky.
    \end{itemize}
    \item Set B (more restrictive)
    \begin{itemize}
        \item Minimum angular separation between the ROI and known sources increased to $3^\circ$, eliminating ROIs where mismodeled sources could bias the background models.
        \item Galactic latitude cut increased to $|b| \ge 30^\circ$, selecting cleaner and more diffuse-dominated parts of the sky.
    \end{itemize}
\end{enumerate}

For each ROI, we use $\Delta \mathrm{BIC}$ and $\Delta \mathrm{AIC}$ to determine, pairwise, which one of two background models is a better description of the 
data from an ROI.  In order to define $\Delta \mathrm{BIC}$ and $\Delta \mathrm{AIC}$, one must determine the number of parameters of the background model.  
For the empirical background models E1 and E2, there are no free parameters.  
But for model FT, there are two parameters associated 
with the diffuse backgrounds, and additional parameters associated with each identified point source.  However, these parameters are determined 
by maximizing the likelihood of the data over the entire $20^\circ \times 20^\circ$ region.  This process does not particularly maximize the 
likelihood in the $1^\circ$ ROI.  As a result, it is not appropriate to consider all of these as free parameters for the purpose of 
defining BIC or AIC.  However, the parameters associated with sources within $3^\circ$ of the ROI 
will noticeably affect the likelihood of the 
data from the ROI.  It is for this reason that the optimization procedure separately floats these parameters again at the last step 
of likelihood maximization, as varying these sources can improve the likelihood of the data in the ROI.  As such, we will treat only 
the parameters associated with sources within $3^\circ$ of the ROI as free parameters for the purpose of defining BIC and AIC.  
In particular, note that for set B, there are initially no sources within $3^\circ$ of the ROI, although new sources in that region 
may be found during the model optimization process.  As a result, we generally expect fewer free parameters for the FT 
models of set B than those of set A.  Since models E1 and E2 in any case have no free parameters, 
for these models we find that 
$\mathrm{BIC}_\mathrm{E1,E2} = \mathrm{AIC}_\mathrm{E1,E2} = -2 \ln {\cal L}_\mathrm{E1,E2}^\mathrm{max}$.

It is important to note that, although we have three approaches to producing a background model in any particular ROI, the 
background models produced by these approaches for any particular ROI are independent of those produced for any other ROI.  It 
is not necessarily true that one approach will tend to produce a background model which better describes the data in every ROI.  
For example, if there is an ROI in which the diffuse astrophysical processes that produce gamma-rays are not well-modeled (for example, 
because the gas content in that region of sky is mismodeled), then a theoretically-motivated background model for that ROI may provide 
a poorer description of the data than a background model obtained purely empirically from off-axis gamma-ray data.  But for another ROI, 
in which the diffuse astrophysical processes are well-modeled and vary slightly with direction, a theoretically-motivated background model 
may provide a better description of the data than a background model obtained empirically from data taken off-axis (where the relevant processes 
are slightly different).  As a result, it is not necessarily meaningful to combine results from many different ROIs, and we will not do so.  
Instead, we determine in what fraction of the ROIs (equivalently, in what fraction of sky away from sources) does one approach produce a background model 
which provides a significantly better description of the data than another approach.

\begin{figure}[t]
    \centering
    \includegraphics[width=0.95\textwidth]{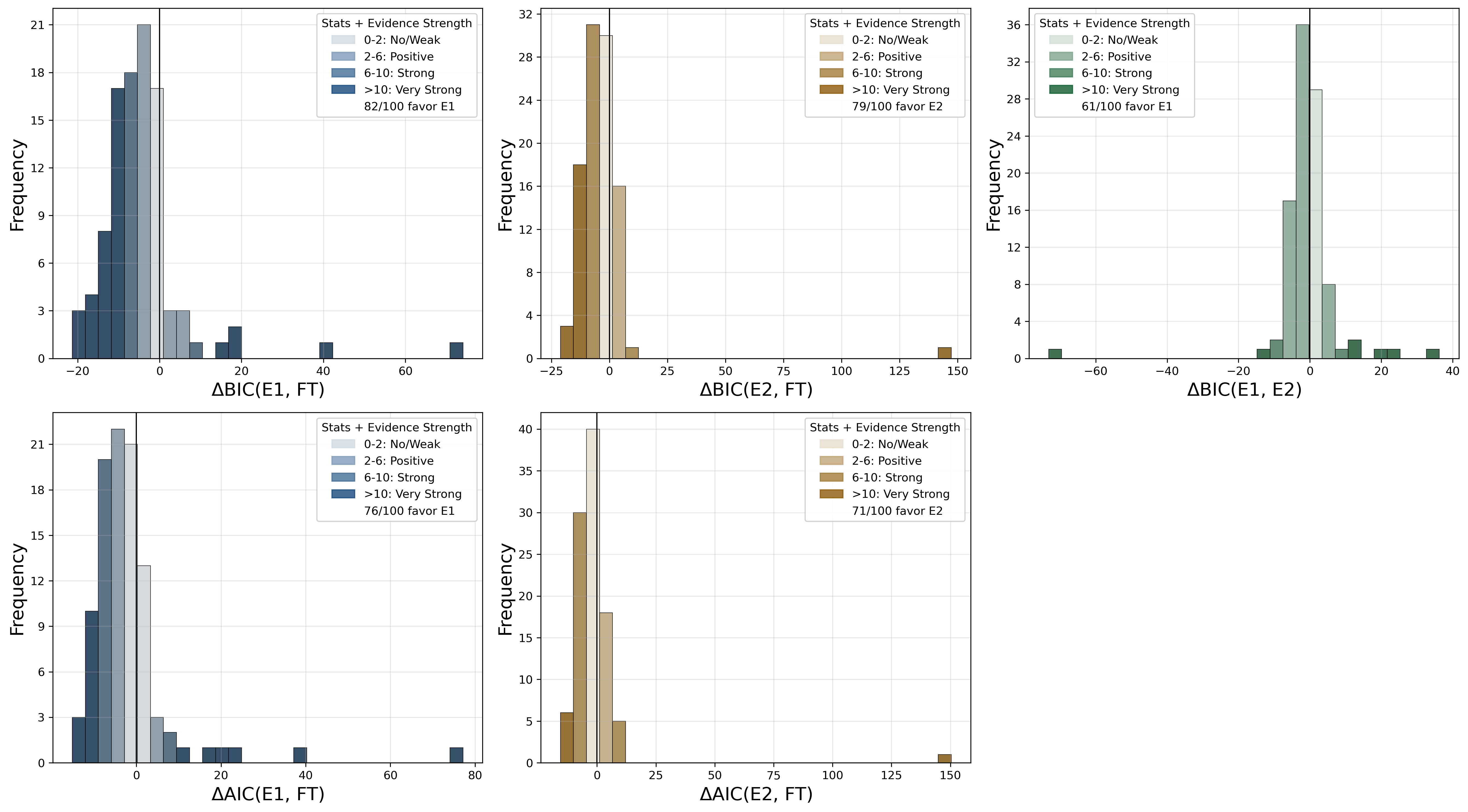}
   \caption{\textbf{Distribution of information-criterion differences for Set A blank-sky regions.} Histograms show the distribution of $\Delta\mathrm{BIC}$ (top row) and $\Delta\mathrm{AIC}$ (bottom row) for the three pairwise model comparisons: $(\mathrm{E1},\mathrm{FT})$, $(\mathrm{E2},\mathrm{FT})$, and $(\mathrm{E1},\mathrm{E2})$. Vertical black lines denote $\Delta \mathrm{B/AIC} = 0$. Shaded regions correspond to conventional evidence-strength thresholds ($|\Delta\mathrm{IC}| = 0$--$2$: weak/no preference; $2$--$6$: positive; $6$--$10$: strong; $>10$: very strong evidence).}
    \label{fig:bicaicSetA}
\end{figure}

\begin{figure}[t]
   \centering
   \includegraphics[width=0.95\textwidth]{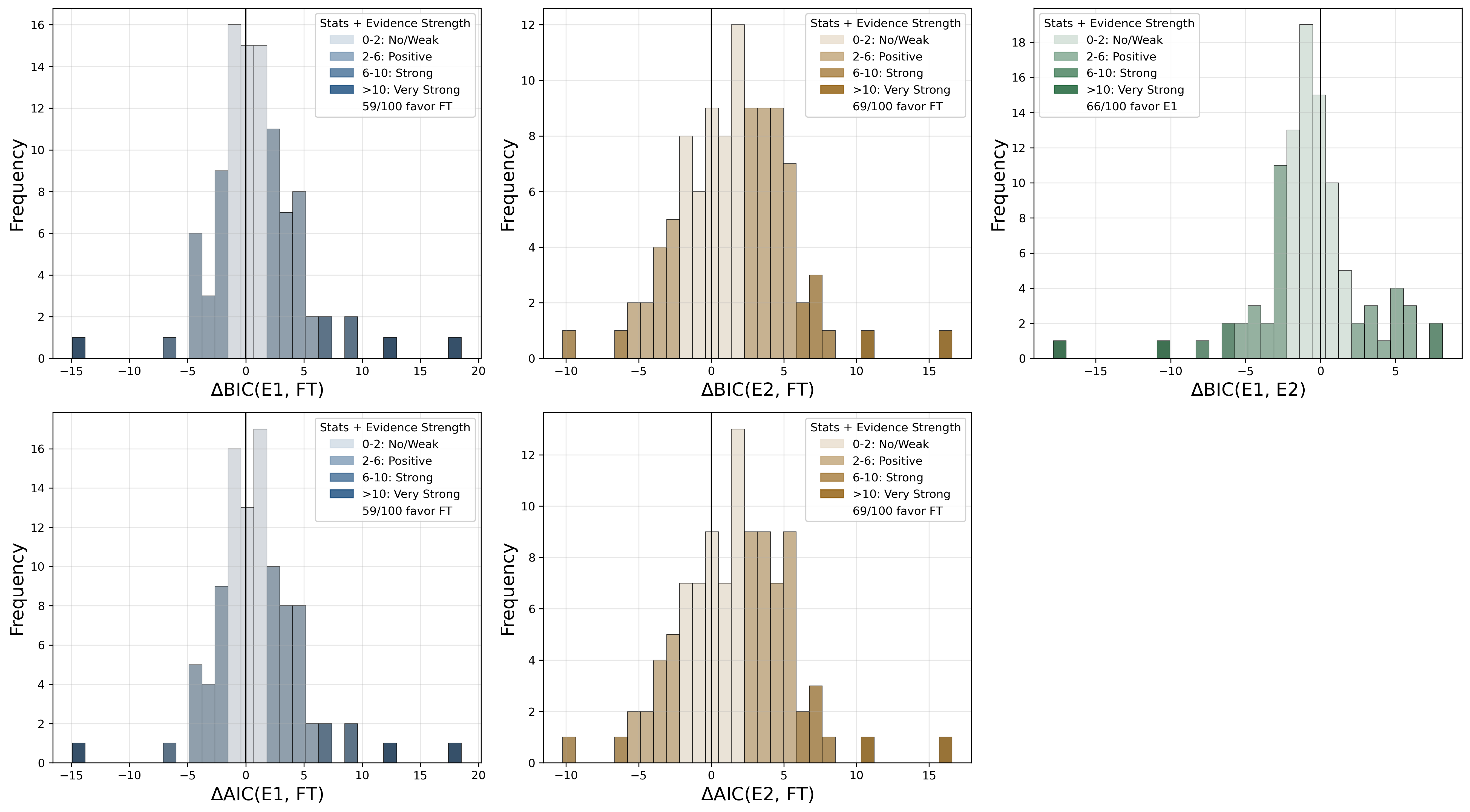}
   \caption{\textbf{Distribution of information-criterion differences for Set B blank-sky regions.} Histograms show the distribution of $\Delta\mathrm{BIC}$ (top row) and $\Delta\mathrm{AIC}$ (bottom row) for the three pairwise model comparisons: $(\mathrm{E1},\mathrm{FT})$, $(\mathrm{E2},\mathrm{FT})$, and $(\mathrm{E1},\mathrm{E2})$. Vertical black lines denote $\Delta \mathrm{B/AIC} = 0$.  Shaded regions correspond to conventional evidence-strength thresholds ($|\Delta\mathrm{IC}| = 0$--$2$: weak/no preference; $2$--$6$: positive; $6$--$10$: strong; $>10$: very strong evidence).}
   \label{fig:bicaicSetB}
\end{figure}

When using $\Delta \mathrm{BIC}$ to compare two models, we will roughly characterize the strength of model preference with the following 
commonly-used standard:

\begin{equation}
    -\Delta \textrm{BIC} \in 
    \begin{cases}
    [0,2): & \textrm{weak,}\\
    [2,6): & \textrm{positive,}\\
    [6,10): & \textrm{strong,}\\
    \geq10: & \textrm{very strong.}
    \end{cases}
\end{equation}
We plot the distributions of $\Delta \mathrm{BIC}$ and $\Delta \mathrm{AIC}$ for the ROIs of sets A and B in Figures~\ref{fig:bicaicSetA} and \ref{fig:bicaicSetB}, respectively.  In the upper three panels of each figure, we plot the distribution of $\Delta \mathrm{BIC}(\mathrm{M1},\mathrm{M2})$, where 
M1 and M2 are chosen pairwise among models E1, E2 and FT, as labeled.  Entries to the left (right) of 0 denote the number (out of 100) 
of ROIs for which model M1 (M2) is preferred.  The darker shaded regions denote values of $\Delta \mathrm{BIC}$ corresponding to a stronger strength of evidence, as labeled.  In each panel, we list the model which provides a better description of the data in most ROIs of the 
set (and how many ROIs that is).
In the lower two panels, we similarly plot the distribution of $\Delta \mathrm{AIC}(\mathrm{M1},\mathrm{M2})$.  
Note, we do not plot the distribution of $\Delta \mathrm{AIC}(\mathrm{E1},\mathrm{E2})$, as $\Delta \mathrm{AIC}(\mathrm{E1},\mathrm{E2}) 
= \Delta \mathrm{BIC}(\mathrm{E1},\mathrm{E2})$.
These results are summarized in Tables~\ref{tab:BIC} and \ref{tab:AIC}.

\subsection{Comparison of models E1 and E2}

We first compare the two empirically-produced background models, E1 and E2.  
We see from Table~\ref{tab:BIC} (right columns) that for sets A and B, 
the preference for either of these empirical models over the other is weak in $36\%$ and $58\%$, respectively, of the 
ROIs.  Although there is some preference for model E1 over model E2 in roughly $2/3$ of the ROIs (see the upper right panels of 
Figures~\ref{fig:bicaicSetA} and~\ref{fig:bicaicSetB}), there is only a strong preference 
for either of the empirically-produced 
models over the other in $18\%$ ($6\%$) of the ROIs of set A (set B).  It thus appears that there is no strong preference 
for one empirically-produced model over the other in most ROIs, even more so when restricting to ROIs in which there 
is less contamination from nearby sources.

It may seem counterintuitive that a background model in which the probability distributions in different energy bins are treated independently is as likely not 
as to better describe the data than a background model in which the covariance between energy bins is explicitly accounted for.  Since a diffuse foreground 
source which may or may not lie directly along the line-of-sight would likely produce photons in multiple energy bins, a background model based on the covariance 
matrix would seem more physically motivated.  However, although model E2 accounts for this covariance, it also assumes that the probability distribution has the form of a multi-variate Gaussian, 
which is only an approximation.  On the other hand, model E1 produces a non-Gaussian PMF  directly from the normalized histogram 
of counts in the sample regions, even though the energy bins are treated independently.  
It could be that one could improve on the multi-variate Gaussian likelihood of model E2 by instead using the multi-variate Gaussian as the kernel 
in approximating the likelihood through kernel density estimation.  However, this would be a computationally intensive process that is beyond the scope of this work.

Since model E1 provides a marginally better description of the data than model E2 
in most ROIs, we can focus the remainder of this discussion on the comparison between models E1 and FT.
Indeed, the comparison between models E2 and FT produces similar results (see Tables~\ref{tab:BIC} and \ref{tab:AIC}).

\subsection{Comparison of models E1 and FT}

As we see in the upper left panel of Figure~\ref{fig:bicaicSetA}, 
for the ROIs in set A, E1 provides a better description of the data than FT in $\sim 80\%$ of the ROIs.  
Moreover, the preference for one model over the other is strong in $\sim 55\%$ of the ROIs, with 
the E1 being strongly preferred in almost all  of these ROIs (see Table~\ref{tab:BIC}).

But as shown in Figure~\ref{fig:bicaicSetB} (upper left panel), for the ROIs in set B, FT provides a better description of the data than the E1 in roughly $60\%$ of the ROIs.  
However, the preference for either of these models is weak in $\sim 50 \%$ 
of the ROIs, and there is only a strong preference for either model in $\sim 10\%$ of the ROIs, with FT being 
strongly preferred in most of these (see Table~\ref{tab:BIC}).

\begin{table}[h]
\centering
\caption{Evidence category counts for pairwise model comparisons using BIC.  For each set of ROIs 
(A or B, as discussed in the text) and for each pair of models (chosen from E1, E2 and FT), entries  
indicate the percent of 100 ROIs in which the given model was preferred, with the listed strength of 
evidence.  When the strength of evidence is weak, no preference for a model is listed.}
\setlength{\tabcolsep}{5pt}
\begin{tabular}{lcccc@{\hspace{1.5em}}cccc@{\hspace{1.5em}}cccc}
\addlinespace[4pt]
\toprule
 & \multicolumn{4}{c}{$\Delta\mathrm{BIC}(\mathrm{E1},\mathrm{FT})$}
 & \multicolumn{4}{c}{$\Delta\mathrm{BIC}(\mathrm{E2},\mathrm{FT})$}
 & \multicolumn{4}{c}{$\Delta\mathrm{BIC}(\mathrm{E1},\mathrm{E2})$} \\
\cmidrule(lr){2-5}\cmidrule(lr){6-9}\cmidrule(lr){10-13}
 & \multicolumn{2}{c}{Set A} & \multicolumn{2}{c}{Set B}
 & \multicolumn{2}{c}{Set A} & \multicolumn{2}{c}{Set B}
 & \multicolumn{2}{c}{Set A} & \multicolumn{2}{c}{Set B} \\
\cmidrule(lr){2-3}\cmidrule(lr){4-5}
\cmidrule(lr){6-7}\cmidrule(lr){8-9}
\cmidrule(lr){10-11}\cmidrule(lr){12-13}
Evidence
 & $\mathrm{E1}$ & $\mathrm{FT}$ & $\mathrm{E1}$ & $\mathrm{FT}$
 & $\mathrm{E2}$ & $\mathrm{FT}$ & $\mathrm{E2}$ & $\mathrm{FT}$
 & $\mathrm{E1}$ & $\mathrm{E2}$ & $\mathrm{E1}$ & $\mathrm{E2}$ \\
\midrule
0--2 (weak)
 & \multicolumn{2}{c}{17\%}
 & \multicolumn{2}{c}{53\%}
 & \multicolumn{2}{c}{14\%}
 & \multicolumn{2}{c}{37\%}
 & \multicolumn{2}{c}{36\%}
 & \multicolumn{2}{c}{58\%} \\
2--6 (positive)
 & 25\% & 3\% & 14\% & 24\%
 & 29\% & 12\% & 15\% & 38\%
 & 28\% & 18\% & 22\% & 14\% \\
6--10 (strong)
 & 24\% & 2\% & 1\% & 5\%
 & 21\% & 2\% & 1\% & 6\%
 & 8\% & 1\% & 2\% & 2\% \\
$>10$ (very strong)
 & 24\% & 5\% & 1\% & 2\%
 & 21\% & 1\% & 1\% & 2\%
 & 3\% & 6\% & 2\% & 0\% \\
\bottomrule
\end{tabular}
\label{tab:BIC}
\end{table}

\begin{table}[h]
\centering
\caption{Evidence category counts for pairwise model comparisons based on AIC, similar to Table~\ref{tab:BIC}. 
Results are shown for Set A only.  For set B, the AIC evidence-category distributions are the same as for BIC 
(see Table~\ref{tab:BIC}). Comparisons between models E1 and E2 are omitted, as 
$\Delta\mathrm{BIC}(\mathrm{E1},\mathrm{E2}) = \Delta\mathrm{AIC}(\mathrm{E1},\mathrm{E2})$ (see Table~\ref{tab:BIC}).}
\setlength{\tabcolsep}{6pt}
\begin{tabular}{lcc@{\hspace{1.2em}}cc}
\addlinespace[3pt]
\toprule
 & \multicolumn{2}{c}{$\Delta\mathrm{AIC}(\mathrm{E1},\mathrm{FT})$}
 & \multicolumn{2}{c}{$\Delta\mathrm{AIC}(\mathrm{E2},\mathrm{FT})$} \\
\cmidrule(lr){2-3}\cmidrule(lr){4-5}
Evidence
 & $\mathrm{E1}$ & $\mathrm{FT}$
 & $\mathrm{E2}$ & $\mathrm{FT}$ \\
\midrule
0--2 (weak)
 & \multicolumn{2}{c}{28\%}
 & \multicolumn{2}{c}{26\%} \\

2--6 (positive)
 & 27\% & 4\%
 & 27\% & 11\% \\

6--10 (strong)
 & 22\% & 3\%
 & 21\% & 8\%\\

$>10$ (very strong)
 & 10\% & 6\%
 & 6\% & 1\% \\

\bottomrule
\end{tabular}
\label{tab:AIC}
\end{table}

It would appear that there is no strong preference for either the FT, E1 or E2 models 
in most ROIs for which there is no contamination from moderately  nearby sources.  However, if one considers ROIs in 
which identified point sources can lie within $3^\circ$ of the ROI, then models E1 and E2 tend to provide descriptions 
of the data that are modestly better than the FT model.  However, it is important to note that this occurs 
because we have selected a background model using BIC, which includes a penalty for additional floating parameters.  
The empirical models have no floating parameters, but if there are sources within $3^\circ$ of the ROI, then their 
normalizations are floated in the final stage of the FT model optimization, and the improvement in the 
maximized likelihood is compensated by the penalty for additional parameters.

Indeed, we can confirm that if one simply compares $2\Delta \log {\cal L}^{\rm max}$ between the FT and E1 
background models, the FT model is preferred in most ROIs within set A ($\sim 80\%$), reflecting the improvement 
in the likelihood which can be obtained from additional parameters.  However, when $\Delta \mathrm{BIC}$ is used as the criterion 
for model selection, we see that the improvement in the likelihood does not justify the floating of the additional parameters.

One would expect that if $\Delta \mathrm{AIC}$ is used as the criterion for model selection, in which there is a weaker penalty for 
additional parameters, then there will be less of a preference for the empirical models among the ROIs of set A.  As we see from the 
lower right panel of Figure~\ref{fig:bicaicSetA}, this 
expectation is borne out.  But even using $\Delta \mathrm{AIC}$, there is a strong preference for one model over the other in 
$\sim 40\%$ of the ROIs of set A, with model E1 preferred over FT in roughly $75\% $ of these ROIs (see Table~\ref{tab:AIC}).

Here, we briefly examine the two most significant outlier ROIs for Set A ($|\Delta\mathrm{BIC}| > 25$, which both favor model FT over E1) to better understand their behavior. For the largest outlier (ROI center: RA = 194.4437$^\circ$, Dec = $-7.01426^\circ$), we note that the bright blazar 3C 279~\cite{Marziani:blz} (4FGL J1256.1$-$0547) lies at an angular separation of $1.289^\circ$ from the ROI center.  

Given the  brightness and variability of 3C 279~\cite{Ballet:2023qzs}, as well as the LAT point-spread function (PSF), 
a non-negligible fraction of photons from 3C 279 may appear in the observed counts from the ROI, but may not be adequately 
modeled by the empirical approaches, in which a mask of radius $0.8^\circ$ is applied around cataloged point sources when creating 
the background distributions.

In contrast, the theoretical modeling framework explicitly incorporates the full spatial and spectral source model (\texttt{PLSuperExpCutoff} spectra) and, therefore, accounts for this emission more consistently.  For this outlier, 
the improvement in the likelihood of the FT model obtained from modeling the nearby source well outweighs the statistical 
penalty due to the extra parameters.

\begin{figure}[t]
    \centering
    \includegraphics[width=0.75\linewidth]{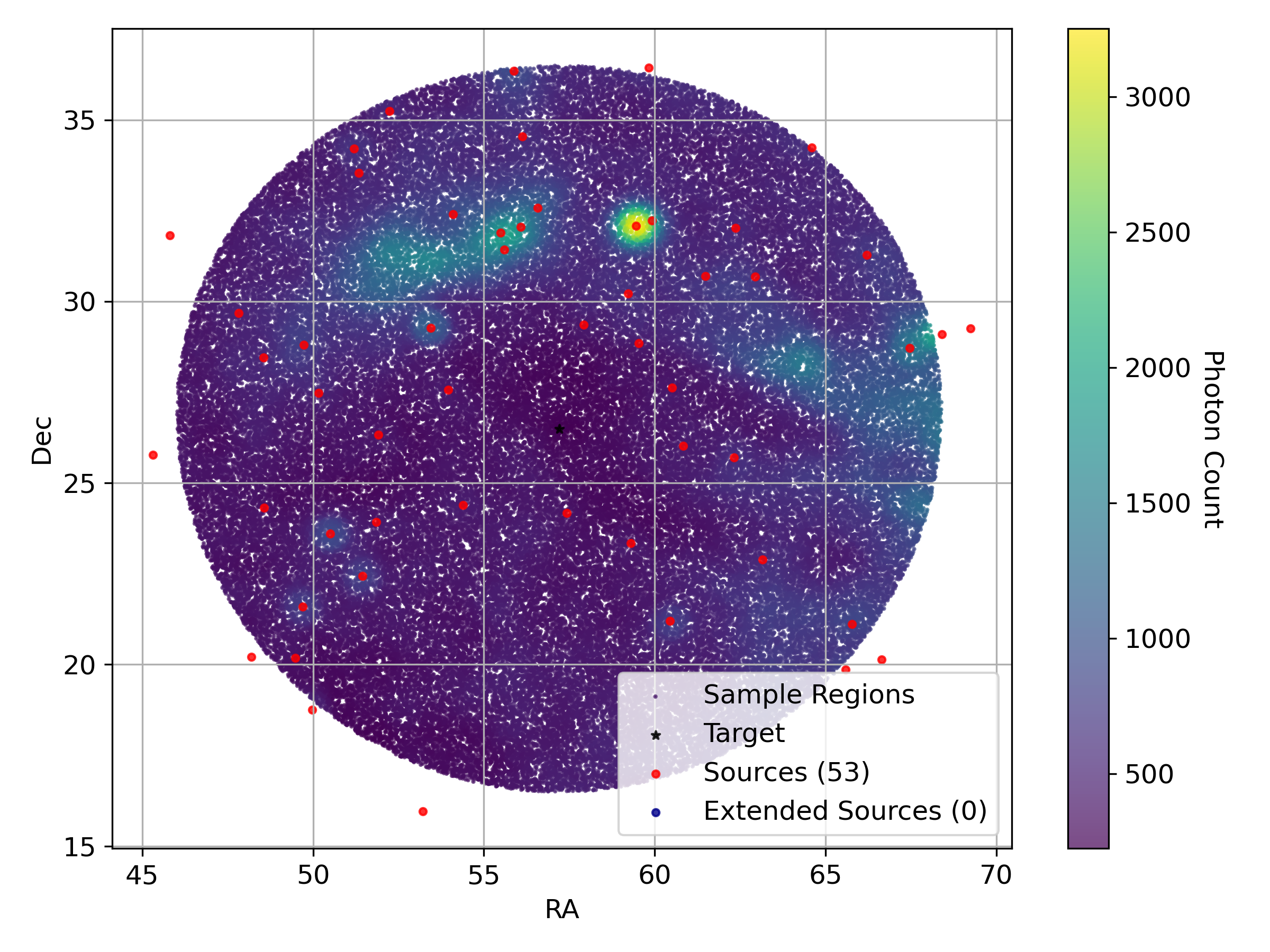}
    \caption{
    Empirical photon count map for ROI center: RA = 57.2065$^\circ$, 
Dec = 26.4928$^\circ$. The map shows the spatial distribution of observed photon counts within the ROI. A coherent extended feature is visible in the upper-left quadrant, which is not captured by the independent empirical model (E1) but is recovered by the covariance and theoretical models. The color scale indicates pure photon counts in the $1-100~\gev$ range.}
    \label{fig:No96_cmap}
\end{figure}

For the second-largest outlier (ROI center: RA = 57.2065$^\circ$, 
Dec = 26.4928$^\circ$), we observe a pronounced spatial feature in the upper-left quadrant of the ROI (see Figure~\ref{fig:No96_cmap}). 
The corresponding BIC values are
$\mathrm{BIC}_{\mathrm{E1}} = 115.72$, 
$\mathrm{BIC}_{\mathrm{E2}} = 80.31$, and 
$\mathrm{BIC}_{\mathrm{FT}} = 76.52$. 
The large $\Delta$BIC/$\Delta$AIC between the E1 and both the E2 and FT models indicates strong evidence against E1 in this region. In contrast, models E2 and FT perform comparably. This structure is visible in the Fermi post-fit model map, and may indicate emission from a diffuse source which produces photons in many energy bins.  Because model E1 treats different energy bins independently, 
it does not capture the impact of these correlations. 
Model E2, however, uses the covariance matrix to model these correlations more effectively.  Model FT uses templates that incorporate the spatial distribution of emission sources, and 
performs even better for this ROI.

Possible improvements to the empirical background modeling could include adapting the masking radius based on source brightness or spectral type (e.g., distinguishing between simple power-law and \texttt{PLSuperExpCutoff} spectra), or dynamically adjusting the mask size according to the LAT energy-dependent PSF. Such refinements may reduce contamination from bright nearby sources.

\section{Conclusion}
\label{sec:Conclusion}

In this work, we applied a likelihood framework to compare empirical and theoretically-derived approaches to modeling gamma-ray backgrounds within the context of a search for dark matter annihilation within a dwarf spheroidal galaxy.  In particular, we have
considered two empirical approaches in which the background probability distribution within a $1^\circ$ region is entirely determined 
from gamma-ray data drawn from regions slightly off-axis.  In one approach (model E1), the probability distribution in each energy bin is assumed 
to be independent of all others, while in the other empirical approach (model E2), the probability distribution is modeled as a 
multi-variate Gaussian with non-trivial covariance.  We have compared these approaches to background modeling using \texttt{Fermitools} (model FT), in which 
the astrophysical processes which can produce gamma-rays in any region of sky are used as an input.

We have compared these different approaches to background modeling by randomly selecting many ``blank sky" regions-of-interest of 
$1^\circ$ opening angle which do not contain any known point sources.  In each region, we have used these three approaches 
to generate a background model for the photon data in the ROI, and computed the likelihood of each background model given the 
observed data.  We have then determined in each ROI which background model provides a better description of the data, using either 
the Bayesian Information Criterion or the Akaike Information Criterion.

We have found that the two empirical approaches, E1 and E2, are equally 
successful.  In most ROIs, the preference for one or the other is not strong.  Similarly, if we consider only ROIs for which there is 
no identified point source within $3^\circ$ of the ROI, then the FT model is as successful as the empirical 
models.  However, if one allows relatively nearby sources, which are modeled with a floating normalization in the FT model, 
then there is a preference for the empirical approach in most of the ROIs.  This appears to be the result 
of the statistical penalty which one must pay for varying parameters in order to maximize the likelihood.  Although the FT models 
tend to have a higher likelihood when nearby sources are modeled with floating parameters, that typically does not 
compensate for the associated statistical penalties (using either BIC or AIC). 
Even so, the preference for the empirical models is only strong in $\sim 40-50\%$ of these ROIs. 
Moreover, there are a few outlier ROIs for which there is a point source which is particularly bright and close to the ROI, 
and for which explicit modeling of the source is worth the 
statistical penalty.

It would be interesting to see if one could provide a better empirical background model by using the multivariate 
Gaussian derived from the covariance matrix as the kernel in a kernel density estimation of the background model.  
It would also be interesting to investigate the impact of modifying the approach to including point source templates 
as part of optimizing the background model.

\section*{Acknowledgments}

We are grateful to Andrea Albert, Natalia Tapia-Arellano, and Shiqi Yu for useful discussions.
JK is supported in part by DOE grant DE-SC0010504.  The work of PS is
supported in part by the National Science Foundation
grant PHY-2412834.
JK and PS wish to acknowledge the Center for Theoretical Underground Physics and Related Areas (CETUP*), the Institute for Underground Science
at Sanford Underground Research Facility (SURF), and the South Dakota Science and Technology Authority for hospitality and financial support,
as well as for providing a stimulating environment.

\bibliography{thebib}

@article{hoskinson:tweedleDEE2024,
  title = {Tools for probing new physics with newly discovered gamma-ray targets},
  author = {Hoskinson, Chance and Kumar, Jason and Sandick, Pearl},
  journal = {Phys. Rev. D},
  eprint = "2408.04611",
  volume = {111},
  issue = {6},
  pages = {063078},
  numpages = {8},
  year = {2025},
  month = {Mar},
  publisher = {American Physical Society},
  doi = {10.1103/PhysRevD.111.063078},
  url = {https://link.aps.org/doi/10.1103/PhysRevD.111.063078}
}

@article{Aramaki:2022zpw,
    author = "Aramaki, Tsuguo and others",
    title = "{Snowmass2021 Cosmic Frontier: The landscape of cosmic-ray and high-energy photon probes of particle dark matter}",
    eprint = "2203.06894",
    archivePrefix = "arXiv",
    primaryClass = "hep-ex",
    journal = "",
    month = "3",
    year = "2022"
}

@article{Boddy:2022knd,
    author = "Boddy, Kimberly K. and others",
    title = "{Snowmass2021 theory frontier white paper: Astrophysical and cosmological probes of dark matter}",
    eprint = "2203.06380",
    archivePrefix = "arXiv",
    primaryClass = "hep-ph",
    reportNumber = "FERMILAB-CONF-22-151-T",
    doi = "10.1016/j.jheap.2022.06.005",
    journal = "JHEAp",
    volume = "35",
    pages = "112--138",
    year = "2022"
}

@article{Strigari:2012acq,
    author = "Strigari, Louis E.",
    title = "{Galactic Searches for Dark Matter}",
    eprint = "1211.7090",
    archivePrefix = "arXiv",
    primaryClass = "astro-ph.CO",
    doi = "10.1016/j.physrep.2013.05.004",
    journal = "Phys. Rept.",
    volume = "531",
    pages = "1--88",
    year = "2013"
}

@article{Boddy:2019kuw,
    author = "Boddy, Kimberly K. and Hill, Stephen and Kumar, Jason and Sandick, Pearl and Shams Es Haghi, Barmak",
    title = "{MADHAT: Model-Agnostic Dark Halo Analysis Tool}",
    eprint = "1910.02890",
    archivePrefix = "arXiv",
    primaryClass = "hep-ph",
    reportNumber = "UH511-1306-2019",
    doi = "10.1016/j.cpc.2020.107815",
    journal = "Comput. Phys. Commun.",
    volume = "261",
    pages = "107815",
    year = "2021"
}

@article{LSST:2008ijt,
    author = "Ivezi\'c, \v{Z}eljko and others",
    collaboration = "LSST",
    title = "{LSST: from Science Drivers to Reference Design and Anticipated Data Products}",
    eprint = "0805.2366",
    archivePrefix = "arXiv",
    primaryClass = "astro-ph",
    reportNumber = "SLAC-PUB-16076",
    doi = "10.3847/1538-4357/ab042c",
    journal = "Astrophys. J.",
    volume = "873",
    number = "2",
    pages = "111",
    year = "2019"
}

@article{LSSTDarkMatterGroup:2019mwo,
    author = "Drlica-Wagner, Alex and others",
    collaboration = "LSST Dark Matter Group",
    title = "{Probing the Fundamental Nature of Dark Matter with the Large Synoptic Survey Telescope}",
    eprint = "1902.01055",
    archivePrefix = "arXiv",
    primaryClass = "astro-ph.CO",
    reportNumber = "FERMILAB-PUB-19-048-A-AE",
    journal = "",
    month = "2",
    year = "2019"
}

@article{Mao:2022fyx,
    author = "Mao, Yao-Yuan and others",
    title = "{Snowmass2021: Vera C. Rubin Observatory as a Flagship Dark Matter Experiment}",
    eprint = "2203.07252",
    archivePrefix = "arXiv",
    primaryClass = "hep-ex",
    reportNumber = "FERMILAB-CONF-22-153-PPD",
    journal = "",
    month = "3",
    year = "2022"
}

@article{Geringer-Sameth:2014qqa,
    author = "Geringer-Sameth, Alex and Koushiappas, Savvas M. and Walker, Matthew G.",
    title = "{Comprehensive search for dark matter annihilation in dwarf galaxies}",
    eprint = "1410.2242",
    archivePrefix = "arXiv",
    primaryClass = "astro-ph.CO",
    doi = "10.1103/PhysRevD.91.083535",
    journal = "Phys. Rev. D",
    volume = "91",
    number = "8",
    pages = "083535",
    year = "2015"
}

@article{Boddy:2018qur,
    author = "Boddy, Kimberly and Kumar, Jason and Marfatia, Danny and Sandick, Pearl",
    title = "{Model-independent constraints on dark matter annihilation in dwarf spheroidal galaxies}",
    eprint = "1802.03826",
    archivePrefix = "arXiv",
    primaryClass = "hep-ph",
    doi = "10.1103/PhysRevD.97.095031",
    journal = "Phys. Rev. D",
    volume = "97",
    number = "9",
    pages = "095031",
    year = "2018"
}

@article{Fermi-LAT:2019yla,
    author = "Abdollahi, S. and others",
    collaboration = "Fermi-LAT",
    title = "{$Fermi$ Large Area Telescope Fourth Source Catalog}",
    eprint = "1902.10045",
    archivePrefix = "arXiv",
    primaryClass = "astro-ph.HE",
    doi = "10.3847/1538-4365/ab6bcb",
    journal = "Astrophys. J. Suppl.",
    volume = "247",
    number = "1",
    pages = "33",
    year = "2020"
}

@article{Ballet:2023qzs,
    author = "Ballet, J. and Bruel, P. and Burnett, T. H. and Lott, B.",
    collaboration = "Fermi-LAT",
    title = "{Fermi Large Area Telescope Fourth Source Catalog Data Release 4 (4FGL-DR4)}",
    eprint = "2307.12546",
    archivePrefix = "arXiv",
    primaryClass = "astro-ph.HE",
    journal = "",
    month = "7",
    year = "2023"
}

@article{Boddy:2024tiu,
    author = "Boddy, Kimberly K. and Carter, Zachary J. and Kumar, Jason and Rufino, Luis and Sandick, Pearl and Tapia-Arellano, Natalia",
    title = "{New dark matter analysis of milky~way dwarf satellite galaxies with madhatv2}",
    eprint = "2401.05327",
    archivePrefix = "arXiv",
    primaryClass = "hep-ph",
    doi = "10.1103/PhysRevD.109.103007",
    journal = "Phys. Rev. D",
    volume = "109",
    number = "10",
    pages = "103007",
    year = "2024"
}

@misc{FermiModel,
      title = {Galactic Interstellar Emission Model for the 4FGL Catalog Analysis},
      author = {{The Fermi-LAT Collaboration }},
      year = {2019},
      note = {\url{ https://fermi.gsfc.nasa.gov/ssc/data/analysis/software/aux/4fgl/Galactic_Diffuse_Emission_Model_for_the_4FGL_Catalog_Analysis.pdf}}
}

@article{Atwood:2013rka,
    author = "Atwood, W. and others",
    collaboration = "Fermi-LAT",
    title = "{Pass 8: Toward the Full Realization of the Fermi-LAT Scientific Potential}",
    eprint = "1303.3514",
    archivePrefix = "arXiv",
    primaryClass = "astro-ph.IM",
    doi = "10.1088/0004-637X/774/1/76",
    journal = "Astrophys. J.",
    volume = "774",
    number = "1",
    pages = "76",
    year = "2013"
}

@misc{FSSC:DataSelection,
    author = "Fermi Science Support Center",
    title = "{LAT Data Selection}",
    howpublished = "Online documentation",
    url = "https://fermi.gsfc.nasa.gov/ssc/data/analysis/scitools/lat_data_selection.html",
}

@article{Geringer_Sameth_2011,
   title={Exclusion of Canonical Weakly Interacting Massive Particles by Joint Analysis of Milky Way Dwarf Galaxies with Data from the Fermi Gamma-Ray Space Telescope},
   volume={107},
   ISSN={1079-7114},
   url={http://dx.doi.org/10.1103/PhysRevLett.107.241303},
   DOI={10.1103/physrevlett.107.241303},
   number={24},
   journal={Physical Review Letters},
   publisher={American Physical Society (APS)},
   author={Geringer-Sameth, Alex and Koushiappas, Savvas M.},
   year={2011},
   month=dec }

@INPROCEEDINGS{Wood:2017vfl,
       author = {{Wood}, M. and {Caputo}, R. and {Charles}, E. and {Di Mauro}, M. and {Magill}, J. and {Perkins}, J.~S. and {Fermi-LAT Collaboration}},
        title = "{Fermipy: An open-source Python package for analysis of Fermi-LAT Data}",
    booktitle = {35th International Cosmic Ray Conference (ICRC2017)},
         year = 2017,
       series = {International Cosmic Ray Conference},
       volume = {301},
        month = jul,
          eid = {824},
        pages = {824},
          doi = {10.22323/1.301.0824},
archivePrefix = {arXiv},
       eprint = {1707.09551},
 primaryClass = {astro-ph.IM},
       adsurl = {https://ui.adsabs.harvard.edu/abs/2017ICRC...35..824W}
}

@misc{Fermitools,
  author = {{Fermi-LAT Collaboration}},
  title = {{Fermitools}: Fermi Science Tools for LAT data analysis},
  url = {https://fermi.gsfc.nasa.gov/ssc/data/analysis/software/},
  year = {2025}
}

@ARTICLE{Marziani:blz,
       author = {{Marziani}, P. and {Sulentic}, J.~W. and {Dultzin-Hacyan}, D. and {Calvani}, M. and {Moles}, M.},
        title = "{Comparative Analysis of the High- and Low-Ionization Lines in the Broad-Line Region of Active Galactic Nuclei}",
      journal = {ApJS},
         year = 1996,
        month = may,
       volume = {104},
        pages = {37},
          doi = {10.1086/192291},
       adsurl = {https://ui.adsabs.harvard.edu/abs/1996ApJS..104...37M}
}

\end{document}